\documentclass{elsart}
\begin{document}                                                               
\input{epsf.tex}

\begin{frontmatter}

\title{Differential Cross Section Measurement \\
		of {\Large $\eta$} Photoproduction on the Proton\\
	 	from Threshold to 1100 MeV}

\author[isn]{F.~Renard},
\author[gen]{M.~Anghinolfi},
\author[rom]{O.~Bartalini},
\author[cat]{V.~Bellini},
\author[isn]{J.P.~Bocquet},
\author[rom]{M.~Capogni},
\author[gen]{M.~Castoldi},
\author[gen]{P.~Corvisiero},
\author[rom]{A.~D'Angelo},
\author[ipn]{J.-P.~Didelez},
\author[rom]{R.~Di~Salvo},
\author[fra]{C.~Gaulard},
\author[tor]{G.~Gervino},
\author[san]{F.~Ghio},
\author[san]{B.~Girolami},
\author[ipn]{M.~Guidal},
\author[ipn]{E.~Hourany},
\author[inr]{V.~Kouznetsov},
\author[ipn]{R.~Kunne},
\author[inr]{A.~Lapik},
\author[fra]{P.Levi~Sandri},
\author[isn]{A.~Lleres},
\author[rom]{D.~Morriciani},
\author[inr]{V.~Nedorezov},
\author[rom,isn]{L.~Nicoletti},
\author[isn]{D.~Rebreyend \thanksref{dom}},
\author[itp]{N.~Rudnev},
\author[gen]{M.~Sanzone}
\author[rom]{C.~Schaerf},
\author[cat]{M.-L.~Sperduto},
\author[cat]{M.-C.~Sutera},
\author[gen]{M.~Taiuti}
\author[kur]{A.~Turinge},
\author[ipn]{Q.~Zhao},
\author[gen]{A.~Zucchiatti}

\collab{GRAAL Collaboration}

\address[isn]{IN2P3, Institut des Sciences Nucl\'eaires, 38026 Grenoble, France} 
\address[gen]{INFN sezione di Genova and Universit\`a di Genova, 16146 Genova, 
		Italy }
\address[rom]{INFN sezione di Roma II and Universit\`a di Roma "Tor Vergata",
		00133 Roma, Italy}
\address[cat]{INFN sezione di Catania and Universit\`a di Catania, 
         	95100 Catania,Italy}
\address[ipn]{IN2P3, Institut de Physique Nucl\'eaire, 91406 Orsay, France}
\address[fra]{INFN Laboratori Nazionali di Frascati, 00044 Frascati, Italy}
\address[tor]{INFN sezione di Torino  and Universit\`a di Torino, 10125 Torino,
		 Italy}
\address[san]{INFN sezione di Roma I and Istituto Superiore di Sanit\`a,  
         	00161 Roma, Italy}
\address[inr]{Institute for Nuclear Research, 117312 Moscow, Russia}
\address[itp]{Institute of Theoretical and Experimental Physics, 117259 Moscow, 
		Russia}
\address[kur]{RRC "Kurchatov Institute", 123182 Moscow, Russia}

\thanks[dom]{E-mail address: rebreyend@isn.in2p3.fr}

\date{\today} 

\maketitle


\begin{abstract}

The differential cross section for the reaction  $p(\gamma, \eta p)$ has been
measured from threshold to 1100~MeV photon laboratory energy. For the first
time, the region of the S$_{11}$(1535) resonance is  fully covered in a
photoproduction experiment and allows a precise extraction of its parameters
at the photon point. Above 1000~MeV, S-wave dominance vanishes while a 
P-wave contribution is observed whose nature will have to be clarified. These
high precision data together with the already measured beam asymmetry data will
provide stringent constraints on the extraction of new couplings of baryon
resonances to the $\eta$ meson.

\vspace{.5cm}

\end{abstract}


\end{frontmatter}


The nucleon, like any composite object, shows a spectrum of excited states 
intimately connected to its internal structure. Precise measurements of the  
properties of these states offer a unique opportunity to test Quantum
ChromoDynamics (QCD)  in the non-perturbative regime and to approach the
confinement problem. Historically, they  have been first observed as baryon
resonances in $\pi$-N scattering \cite{vra00}, with a  rich spectrum typical of
a few-body problem. Since the dominant decay  channel of nucleon resonances is
the strong decay through meson emission, photoproduction  of light mesons
($\pi$, $\eta$, K,...) gives a complementary way to access information  about
nucleon spectroscopy. Whereas elastic and inelastic $\pi$-N scattering have
provided  precise values for the masses and widths \cite{pdg00}, meson
photoproduction allows a measurement of the electromagnetic transitions, thus 
provides a stringent dynamical test for models of the nucleon.  

Among the large variety of exis\-ting mo\-dels \cite{bha88}, Constituent Quark
Models (CQM) have been the most successful in accounting for the observed
spectrum. These  "QCD-inspired" models describe the nucleon as three massive
constituent quarks  ($m_Q\approx$300~MeV$/c^2$) confined by a harmonic
potential. The hyperfine interaction,  essential to reproduce the spectrum, is
derived from gluon exchange in the original approach,  or meson exchange in the
more recent chiral version of CQM \cite{vra00}. Both versions also  predict so
far unobserved  states ("missing resonances") whose absence has been
interpreted, for instance, by a weak  coupling to $\pi$-N channels \cite{cap94}. 

Extraction of resonance properties in pion photoproduction is a difficult
exercise since one needs to disentangle many overlapping contributions
\cite{said,maid}.  By contrast, eta photoproduction close to threshold is
strongly dominated by a single resonance, the S$_{11}$(1535), and thus is the
ideal place to investigate this benchmark state for nucleon models.  Analyses
of the available data base, including differential cross sections below 800~MeV
\cite{kru95}, target asymmetries below 1000~MeV \cite{boc98} and our beam 
asymmetry $\Sigma$ measurements \cite{aja98} have identified two other
resonances : the  D$_{13}$(1520) and the F$_{15}$(1680).  On the one hand,
$\Sigma$ was shown to be very sensitive to the D$_{13}$ and allowed a precise
extraction of its parameters \cite{zli98,muk98,tia99,wor99}. On the other hand,
the F$_{15}$(1680) was suggested to explain the forward peaking observed in 
$\Sigma$ above 1000~MeV, but, in the absence of cross section data at this
energy, no definite conclusion could be reached. 

In this paper, we report on measurements of the differential cross section and
extraction of the total cross section   for the reaction $p(\gamma, \eta p)$
from threshold E$_{\gamma}$=707~MeV (W=1485~MeV)  to 1100~MeV 
(W=1716~MeV), where W is the total CM energy.

These data have been obtained with the GRAAL facility, located at the ESRF
(European Synchrotron  Radiation Facility) in Grenoble. The polarized and
tagged photon beam is created by Compton  backscattering of laser light from
the 6.04 GeV electrons circulating in the storage ring. The measurements
presented here used the green line ($\lambda$=514~nm) of an Argon laser;   the
$\gamma$-ray energy spectrum extends from 500~MeV (geometrical limit of the
tagging system)  to the Compton edge at 1100~MeV. Detailed description of the
beam, 4$\pi$ detector characteristics  and acquisition system can be found in
refs. \cite{aja98,lev96,bar97,fre99,aja00}. An energy deposition in the BGO
calorimeter larger than 180~MeV  in coincidence with an electron in the tagging
system, triggers the data acquisition and allows simultaneous recording of the
$\pi^0 p$ and $\eta p$ channels.

\vspace*{.2cm} 
\begin{figure} 

\centerline{\epsfverbosetrue\epsfxsize=8.cm\epsfysize=4.8cm
\epsfbox{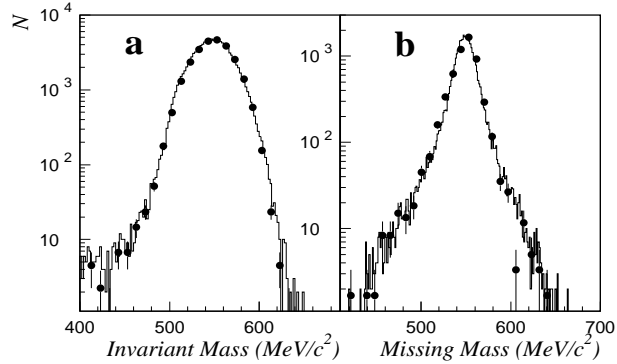}} 

\caption{\label{Figure1} a: Invariant mass spectrum for $\eta\!\rightarrow\!
2\gamma$ decay.
b: Missing mass spectrum, as calculated from the proton momentum.  Data 
(full curve) and simulation  (black dots) are  presented with all  
kinematical cuts applied.} 
\vspace*{-.1cm} 
\end{figure} 

Complete detection of the reaction products is required in the event analysis.
The photons from neutral $\eta$ decays ($\eta\!\rightarrow\!2\gamma$ and
$\eta\!\rightarrow\!3\pi^0\!\rightarrow\!6\gamma$) are  identified in the BGO
calorimeter, while the recoil proton is tracked in wire chambers and
characterized by time of flight (ToF)  and dE/dx measurements. With the tagger
providing the energy of the incoming photon, the  kinematics is overdetermined
and a clean event selection is easily achieved using kinematical  cuts (6 in
total). Two examples are given with the invariant mass of the $\eta$ (Fig.~1-a)
and the missing mass calculated from the proton momentum (Fig.~1-b). For both
quantities, the simulated spectrum nicely reproduces  the experimental shape,
which indicates the reliability of the simulation program and the absence of
any  electromagnetic background. The average hadronic background has been
estimated to be less than  1\% \cite{fre99}.

The cross section normalization takes into account the target thickness, the
photon beam  intensity, the detection efficiency and the branching ratios of
the $\eta$-meson decays taken  from \cite{pdg00} ($\eta\!\rightarrow\!
2\gamma$: 39.21$\pm$0.34~\%, $\eta\!  \rightarrow\! 3 \pi^0$:
32.2$\pm$0.4~\%).  The thickness of the  liquid hydrogen target is
0.217$\pm$0.003~g$\cdot$cm$^{-2}$, which gives a contribution  of 1.5\% to the
systematic errors. Empty target runs have indicated a contribution of the
target  walls of 0.9\%, consistent with their thickness. The photon intensity
is monitored by thin  plastic scintillators located between the target and a
total absorption calorimeter (Spacal)  which serves as a beam dump. Both
detectors are in coincidence with the tagging system and  their ToF spectra are
measured for correction of accidentals. The low efficiency of the thin  monitor
($\simeq$~2.7\%)  prevents pile-up effects at high photon rate and is measured
by  comparison with the Spacal rate at low flux.


\begin{figure}

\centerline{\epsfverbosetrue\epsfxsize=9.cm\epsfysize=6.cm
\epsfbox{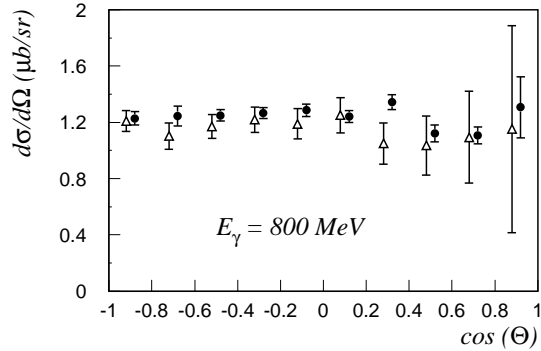}}

\caption{\label{Figure2}
Differential cross section for the $p(\gamma, \eta p)$ reaction at 800~MeV. 
Results for events with $\eta\!\rightarrow\!2\gamma$ (black dot) 
and $\eta\!\rightarrow\!3\pi^0$ (open triangles) decays are compared.}
\end{figure}

The detector efficiency (50\% on average) is calculated with a complete 
Monte-Carlo simulation of the apparatus, including the dependence of beam shape
on energy and polarization, using a realistic event generator \cite{maz94}.
Apart from the  geometrical acceptance, the main loss comes from the overlap of
clusters associated with  $\gamma$-rays in the BGO ball (4 crystals hit on
average per photon). This effect is strongly correlated to the  $\gamma$-ray
multiplicity as well as the cluster threshold, and is, moreover, energy
dependent. In order to test the reliability of our simulation code,  the
differential cross section has been calculated simultaneously for the two
$\eta$ decays:  $\eta\rightarrow 2\gamma$ and  $\eta\rightarrow 3\pi^0$. The
comparison at one energy  (800~MeV) is displayed in Fig.~2 and illustrates the
excellent agreement between both results. In addition, we have systematically
used the high statistics reaction  $p(\gamma, \pi^0 p)$ to perform precise
tests and comparisons with the world data  base \cite{fre99}. The error bars
shown in the data correspond to the quadratic sum of  all systematic and
statistical errors except for those due to global normalisation factors 
estimated to be $\pm$3.0\%.

The total cross section has been obtained by integration of the differential
cross section, using a  polynomial fit in $\cos \theta$  to extrapolate to the
unmeasured region ($\simeq$10\%). A good reduced $\chi^2$ is already achieved
with a polynomial of degree two in agreement with  the smooth behavior observed
in our measurements. It should be  noticed that this extrapolation procedure is
a source of error that cannot be evaluated experimentally.

The total cross section calculated  from threshold to 1100~MeV is plotted    
in Fig.~3. One can notice the nice agreement  with previous measurements 
obtained at Mainz \cite{kru95} except for a small discrepancy around 800~MeV. 

\vspace*{.5cm}
\begin{figure}

\centerline{\epsfverbosetrue\epsfxsize=8.0cm\epsfysize=6.cm
\epsfbox{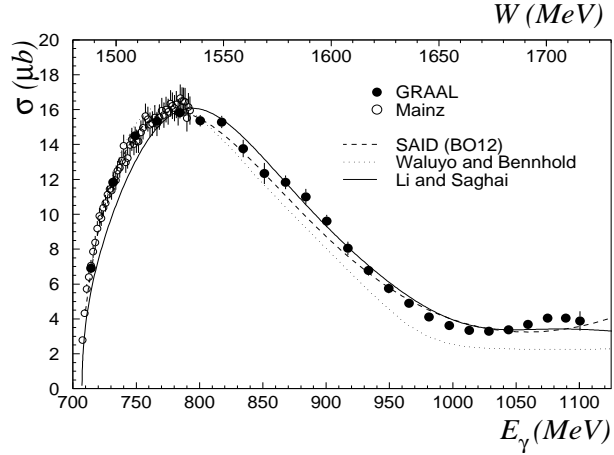}}
\vspace*{.1cm}
 
\caption{\label{Figure3}Total cross section of the 
$p(\gamma, \eta p)$ reaction. The GRAAL results (close circles) are 
compared with previous experimental results and to three different analyses. 
See text for details.}
\end{figure}

As mentioned earlier, the S$_{11}$(1535)  resonance strongly dominates $\eta$
photoproduction up to 900~MeV and its parameters  (mass, width and
photocoupling amplitude) can be directly estimated from the shape of the total
cross section. A fit with a single Breit-Wigner shape limited to energies less
than 900~MeV gives M=1543$\pm$2~MeV and $\Gamma_R$=174$\pm 8$~MeV for the mass
and width of the resonance. Extending the fit to higher energies results in a
much smaller width (for instance, $\Gamma_R = 152 \pm 4$~MeV for E$_{\gamma}
\leq 950$~MeV) and underlines the presence of other contributions. Our
extracted $\Gamma_R$ is significantly smaller than the one obtained with the
Mainz data ($\geq$200~MeV) \cite{kru95} but is in agreement with the PDG
estimate (150~MeV) and also with two recent electroproduction experiments
performed at JLab  ($\Gamma_R \approx 150$~MeV) \cite{arm99,tho01}. The
helicity amplitude A$_{1/2}$ can be readily obtained from the maximum of the
cross section. It shows less sensitivity to the energy range covered  but is
correlated to the branching ratio of the resonance to the $\eta p$ channel,
b$_{\eta}$. Taking  b$_{\eta}$=0.55 for consistency with extractions from the
two JLab electroproduction experiments, we obtain
A$_{1/2}$=102$\pm$2~10$^{-3}$~GeV$^{-1/2}$. This value falls within the broad
range of the PDG estimate ($ 90 \pm 30$), but, taking into account other
contributions, especially the second S$_{11}$(1650), might significantly
change the result.

\begin{figure}

\centerline{\epsfverbosetrue\epsfxsize=11.cm\epsfysize=9.cm
\epsfbox{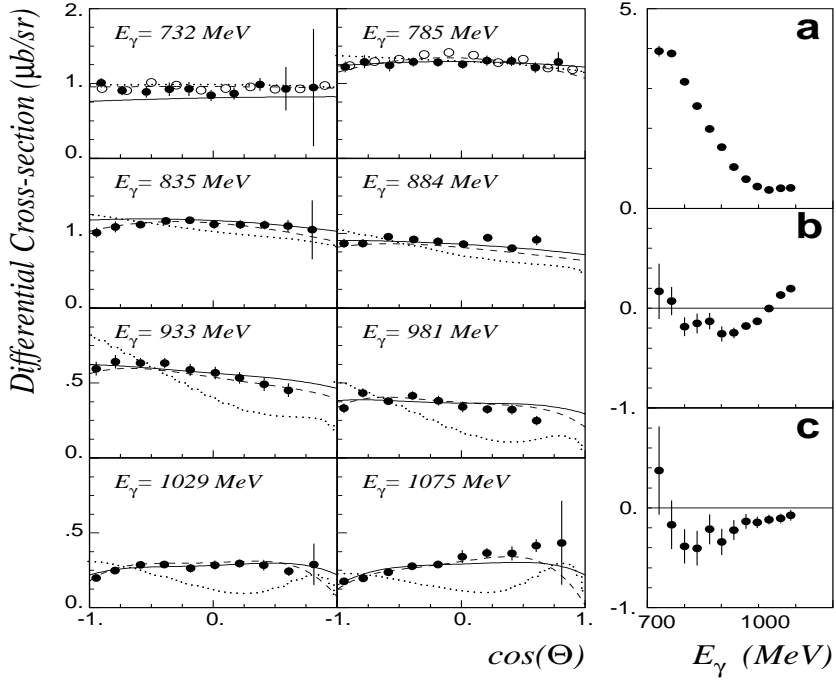}}

\caption{\label{Figure4} Differential cross section for the $p(\gamma, \eta p)$
reaction. Legend as in Fig.~3. The coefficients a, b and c result from the fit 
to the data by expression (1).}
\end{figure}

The differential cross section, plotted in Fig.~4 for a sample of photon
energies as a function  of $\cos \theta$ where  $\theta$ is the $\eta$ C.M.
polar angle, was measured every 17~MeV between threshold  and 1100~MeV 
(233 points). For the two lowest energies of the figure, data from Mainz are
also shown and illustrate again the good agreement between both experiments. 
Thanks to the overwhelming dominance of the S$_{11}$(1535) resonance,  the
differential cross section can be expanded in terms of the S-wave multipole and
its interference  with other multipoles. Limiting the expansion to P and D
waves, an approximation valid  at least up to 900~MeV, the following expression
can be derived~\cite{tia99}: 

\begin{center}
$\frac{d\sigma}{d\Omega}=\frac{q_{\eta}}{k}\{a + b\, \cos\theta+ c\, 
\cos^2\theta\}\quad (1) $ 
\end{center}

where $q_{\eta}$ and $k$ are the $\eta$ and $\gamma$ C.M. momenta, and 

\hspace{1cm}$a=|E_{0^+}|^2-Re[E_{0^+}^*(E_{2^-}-3M_{2^-})]$

\hspace{1cm}$b=2\, Re[E_{0^+}^*(3E_{1^+}+M_{1^+}-M_{1^-})]$

\hspace{1cm}$c=3\, Re[E_{0^+}^*(E_{2^-}-3M_{2^-})]$

 We have fitted our results with expression (1) and the  extracted coefficients
are plotted versus $\gamma$-ray energy on the right-hand side of Fig.~4.   As
expected, the D$_{13}$(1520) contribution already seen in ref. \cite{kru95}
close to threshold is observed in c (S-D interference). The b coefficient (S-P
interference) exhibits small and negative values below 1000~MeV and then a
sudden rise with a zero-crossing at E$_{\gamma}$=1030~MeV (W=1680~MeV), which
indicates the onset of a P-wave. A similar trend has been observed in
electroproduction at low Q$^2$ \cite{tho01} where this P-wave has been
associated with the P$_{11}$(1710) resonance. Above 1000~MeV, nevertheless, a
simple interpretation  of these coefficients becomes doubtful since S-wave
dominance vanishes. 

A multipole analysis has been performed by the GWU group \cite{str00}. They
achieved a nice global fit to the cross section  (dashed line, Figs.~3 and 4)
and asymmetry (Fig.~5); for the latter, we have displayed the highest
energy bin published in 1998 and a new measurement using a UV laser line that
confirms the large values at forward angle. Only S, P and D multipoles are
needed; this rules out a strong contribution from the F$_{15}$(1680).
Analysis of the individual multipoles confirms the previous qualitative
conclusions and indicates strong P and D-wave contributions above 900~MeV. 
These higher multipoles can indeed be associated with nucleon resonances
(P$_{11}$(1710) or P$_{13}$(1720), D$_{15}$(1675) or D$_{13}$(1700)), but
additional contributions can originate from vector meson exchange as well as
nucleon Born terms.  In order to start exploring this new domain, we present
below preliminary results of two models which  bring  different answers to
this question.

\begin{figure}

\centerline{\epsfverbosetrue\epsfxsize=6.cm\epsfbox{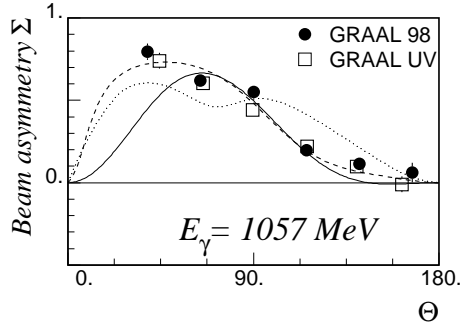}}
\label{Figure5} 
\vspace*{.1cm}
\caption{Beam asymmetry $\Sigma$ for the $p(\gamma, \eta p)$ reaction. Results 
published in 1998 (circles) are compared to new results obtained with a UV 
line (squares). Legend as in Fig.~3} 
\end{figure}

The first, from Waluyo and Bennhold, \cite{wal00,ben99} is a coupled-channel
calculation  based on the Bethe-Salpeter equation in the K-matrix
approximation. It includes all hadronic and  electromagnetic reactions with
$\gamma$N, $\pi$N, $\pi$$\pi$N, $\eta$N, K$\Lambda$, K$\Sigma$ and $\eta'$N
asymptotic states. Nucleon resonances with  spin J $\leq$ 3/2 are included up
to M~=~2~GeV. The results, based on a large data base, are plotted in Figs. 3
to 5 (dotted line). They show fair overall agreement, especially for $\Sigma$, 
and to a lesser extent for $d\sigma$/$d\Omega$, in particular at high energy.
Their unexpected conclusion is that above 900~MeV, the forward peaking of
$\Sigma$  is primarily due to intermediate vector meson exchange in the {\it
t}-channel.

The second, from Li and Saghai, \cite{zli98,sag00} is based on a quark model
and includes all known resonances. To avoid double counting, vector meson
exchange in the {\it t}-channel is excluded, a valid approximation as long as
resonant contributions dominate. An interesting property  of this approach is
that it directly links  data to the quark model, hence to quantities such as
mixing angles of resonance configurations \cite{sag01}. The model, constrained
by our data only, is able to reproduce rather well the global trend for the
cross section (full curve), although significant deviations are observed  near
threshold and above 1000~MeV. The shape of the beam asymmetry is also fairly
well reproduced but, despite its contribution, the F$_{15}$(1680) cannot 
explain the forward peaking.  In their most recent analysis \cite{sag02}, the
authors have shown that, thanks to the inclusion of a new S$_{11}$ resonance
around 1700~MeV, they obtain a significant improvement and are now able to
reproduce nicely the set of data, except the forward asymmetry at 40$^0$ and
1057~MeV.

In summary, we have measured the differential cross section for the reaction 
$p(\gamma, \eta p)$ from threshold to 1100~MeV photon lab. energy, completing
our previous measurement of the beam asymmetry $\Sigma$ over the same energy
range. Below 900~MeV, the reaction mechanism is well understood and these data
will contribute to a precise determination of the dominant S$_{11}$(1535)
parameters. Above this energy, S-wave dominance vanishes and the data show a
rapid transition to a new regime with a clear P-wave contribution. It seems
established that the F$_{15}$(1680) resonance is not the main source of the
forward peaking observed  in $\Sigma$. We are looking forward to definitive
analyses that will allow the extraction of as yet unknown couplings of  baryon
resonances to the $\eta$ meson. 

\ack

We are grateful to C. Bennhold, B. Saghai and I. Strakovsky for helpful
discussions and communication of their analyses prior to publication. It is a
pleasure to thank the ESRF for the reliable and stable operation of the storage
ring  and the technical staff of the contributing institutions for essential
help in the realization  and maintenance of the apparatus.




\begin{thebibliography}{34}

\bibitem{vra00} For a recent review, see T.P.~Vrana, S.A.~Dytman, T.S.H.~Lee, 
		Phys. Rep. {\bf 328} (2000) and references therein.
\bibitem{pdg00} Review of Part. Phys., Eur. Phys. J. {\bf C15}, 1-878 (2000). 
\bibitem{bha88} Models of the nucleon, R.K. Bhaduri (Addison Wesley, 1988).
\bibitem{cap94} S. Capstick and W. Roberts, Phys. Rev. {\bf D47}, 1994 (1993); 
			Phys. Rev. {\bf D49}, 4570 (1994).
\bibitem{said} R.A. Arndt, I.I. Strakovsky and R.L. Workman, Phys. Rev. 
		{\bf C56}, 577 (1997).
\bibitem{maid} D. Drechsel, O. Hanstein, S.S. Kamalov and L. Tiator, Nucl. 
		Phys. {\bf A645}, 145 (1999).
\bibitem{kru95} B. Krusche {\it et al.}, Phys. Rev. Lett. {\bf 74}, 3736 (1995).
\bibitem{boc98} A. Bock {\it et al.}, Phys. Rev. Lett. {\bf 81}, 534 (1998).
\bibitem{aja98} J. Ajaka {\it et al.}, Phys. Rev. Lett. {\bf 81}, 1797 (1998).
\bibitem{zli98} Z. Li and B. Saghai, Nucl. Phys. {\bf A644}, 345 (1998).
\bibitem{muk98} N. Mukhopahyay and N. Mathur, Phys. Lett. {\bf B444}, 7 (1998).
\bibitem{tia99} L. Tiator, D. Drechsel, G. Knochlein, C. Bennhold, Phys. Rev.
		{\bf C60}, 035210 (1999).
\bibitem{wor99} R. Workman, R.A. Arndt, I. Strakovsky, Phys. Rev. {\bf C62}, 
		048201 (2000).
\bibitem{lev96} P. Levi Sandri {\it et al.}, Nucl. Inst. Meth. {\bf A370}, 396
		(1996).
\bibitem{bar97} D. Barancourt {\it et al.}, Nucl. Inst. Meth. {\bf A388}, 226
		(1997).
\bibitem{fre99} F. Renard, Thesis, Univ. J. Fourier (Grenoble, 1999), available
		upon request.
\bibitem{aja00} J. Ajaka {\it et al.}, Phys. Lett. {\bf B475}, 372 (2000).
\bibitem{maz94} L. Mazzaschi {\it et al.}, Nucl. Inst. Meth. {\bf A436}, 441 
		(1994).
\bibitem{arm99} C.S. Armstrong {\it et al.}, Phys. Rev. {\bf D60}, 052004 
		(1999).	     
\bibitem{tho01} R. Thompson {\it et al.}, Phys. Rev. Lett. {\bf 86}, 1702 
		(2001).	     
\bibitem{str00} R.A. Arndt, I.I. Strakovsky and R.L. Workman, BAPS {\bf 45}, 
		43 (2000) and  private communication.
\bibitem{wal00} A. Waluyo and C. Bennhold, private communication.
\bibitem{ben99} C. Bennhold {\it et al.}, nucl-th/9901066, nucl-th/0008024,
			T.~Feuster and U.~Mosel, Phys. Rev. {\bf C59}, 460 (1999).
\bibitem{sag00} B. Saghai, private communication.
\bibitem{sag01} B. Saghai, N*2000 Workshop Proceedings (to appear in World 
		Scientific), JLab (2000).
\bibitem{sag02} B. Saghai and Z. Li, Eur. Phys. J. A {\bf 11},217 (2001).


\end{thebibliography}
\end{document}